\documentstyle[aps,epsf,rotate,subfigure]{revtex}
\font\mb=msbm10
\begin{document}
\draft
\title{Transitions from deterministic to stochastic diffusion}
\author{R. Klages\cite{em}}
\address{Max Planck Institute for Physics of Complex Systems,
N\"othnitzer Str. 38,  D-01187 Dresden, Germany} 
\date{\today}
\maketitle
\begin{abstract}
We examine characteristic properties of deterministic and stochastic diffusion
in low-dimensional chaotic dynamical systems. As an example, we consider a
periodic array of scatterers defined by a simple chaotic map on the
line. Adding different types of time-dependent noise to this model we compute
the diffusion coefficient from simulations. We find that there is a crossover
from deterministic to stochastic diffusion under variation of the perturbation
strength related to different asymptotic laws for the diffusion
coefficient. Typical signatures of this scenario are suppression and
enhancement of normal diffusion. Our results are explained by a simple
theoretical approximation.
\end{abstract}
\pacs{PACS numbers:  05.60.Cd, 05.45.Ac, 05.40.Jc}
To understand diffusion in {\em noisy maps}, that is, in time-discrete
dynamical systems where the deterministic equations of motion are perturbed by
noise, figures as a prominent problem in recent literature. The most simple
example of such models are one-dimensional chaotic maps on the line. In
seminal contributions by Geisel and Nierwetberg \cite{GeNi82}, and by Reimann
et al.\ \cite{Rei94}, scaling laws have been derived for the diffusion
coefficient yielding suppression and enhancement of diffusion with respect to
variation of the noise strength. Related results have been obtained in Refs.\
\cite{Fra91,Wack99}. However, all these results apply only to the onset of
diffusion where the scaling laws are reminiscent of a dynamical phase
transition, and not much appears to be known far away from this transition
point. In such more general situations, only perturbations by a nonzero
average bias have been studied \cite{BaKl97}. Related models are deterministic
Langevin equations, in which the interplay between deterministic and
stochastic chaos has been analyzed \cite{BeRo87}, however, without focusing on
diffusion coefficients. Non-diffusive noisy maps have furthermore been
investigated by refinements of cycle expansion methods \cite{Cvit00}.

In this work we study the transition scenario from deterministic to stochastic
diffusion in the most simple type of chaotic dynamical systems, which are
piecewise linear maps on the line. Particularly, we are searching for
signatures of deterministic and stochastic dynamics in the diffusion
coefficient as a function of the strength of time-dependent stochastic noise.
In this aspect our work appears to be related to the recent dispute on a
possible distinction between chaotic and stochastic diffusion in experiments
\cite{GBFS+98}, where some of the theoretical models studied are very similar
to the one introduced below.

\begin{figure}[b]
\epsfxsize=12cm
\centerline{\rotate[r]{\epsfbox{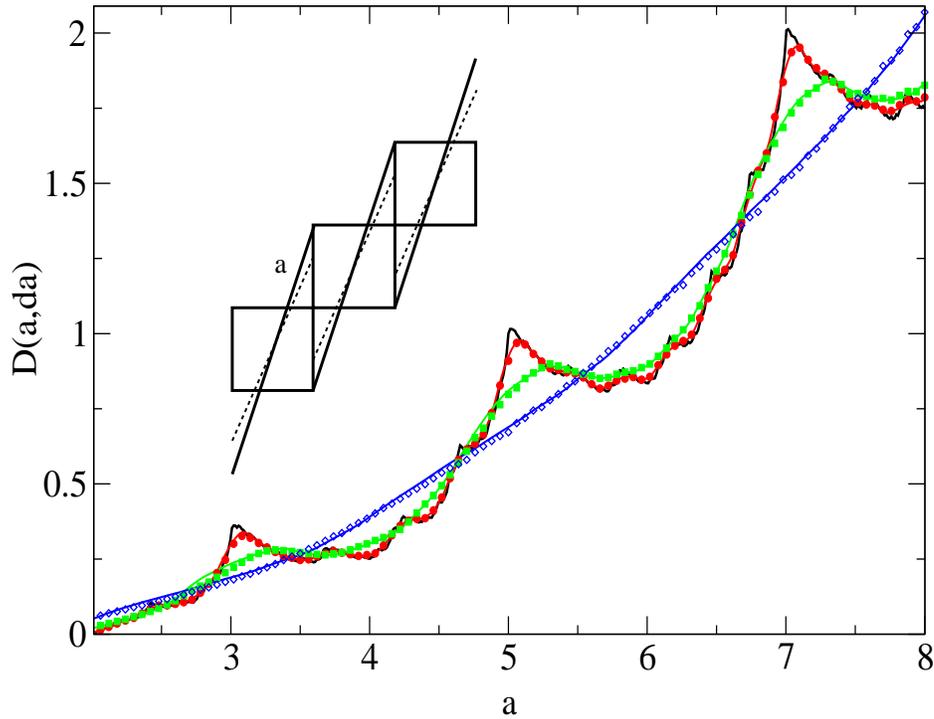}}}
\caption{Diffusion coefficient $D(a,da)$ for the piecewise linear map shown in the
figure. The slope $a$ is perturbed by uniform noise of maximum strength $da$
as defined in Eq.\ (\ref{eq:urp}). The bold black line depicts numerically
exact results for the unperturbed diffusion coefficient at $da=0$. Computer
simulation results for $da\neq0$ are marked with symbols, the corresponding
lines are obtained from the approximation Eq.\ (\ref{eq:dappg}).  The
parameter values are: $da=0.1$ (circles), $da=0.4$ (squares), $da=1.0$
(diamonds).}
\label{fig1}
\end{figure}

We define our system as follows: The unperturbed map is given by the equation
of motion
\begin{equation}
x_{n+1}=M_a(x_n) \quad , \label{eq:eom}
\end{equation}
where $a\in \hbox{\mb R}$ is a control parameter and $x_n$ is the position of
a point particle at discrete time $n$. $M_a(x)$ is continued periodically
beyond the interval $[-1/2,1/2)$ onto the real line by a lift of degree one,
$M_a(x+1)=M_a(x)+1$. We assume that $M_a(x)$ is anti-symmetric with respect to
$x=0$, $M_a(x)=-M_a(-x)$. The map we study as an example is defined by
$M_a(x)=ax$, where the uniform slope $a$ serves as a control parameter. The
Lyapunov exponent of this map is given by $\lambda=\ln a$ implying that for
$a>1$ the dynamics is chaotic. We now apply two types of annealed disorder to
this map, (i) noisy slopes \cite{Rei94,Wack99}: we add the random variable
$\Delta a_n$, $n\in \hbox{\mb N}$, to all slopes $a$ making them
time-dependent in form of
\begin{equation}
M_{a+\Delta a_n}(x)=(a+\Delta a_n)x\;, \label{eq:rs}
\end{equation}
or (ii) noisy shifts \cite{GeNi82,Rei94,Fra91}: we add the random
variable $\Delta b_n$, $n\in \hbox{\mb N}$, as a time-dependent uniform bias
yielding
\begin{equation}
M_{a,\Delta b}(x)=ax+\Delta b_n\;.
\end{equation}
In both cases we assume that the random variable $\Delta_n\in\left\{\Delta
a_n,\Delta b_n\right\}$ is independent and identically distributed according
to a distribution $\chi_{d}(\Delta_n)$, where $d\in\left\{da,db\right\}$ is
again a control parameter. In the following we will consider two different
types of such distributions, namely random variables distributed uniformly
over an interval of size $[-d,d]$ \cite{Fra91,Wack99},
\begin{equation}
\chi_{d}(\Delta_n)=\frac{1}{2d}\Theta(d+\Delta_n)\Theta(d-\Delta_n) \quad , \label{eq:urp}
\end{equation}
and dichotomous or $\delta$-distributed random variables \cite{Rei94,Wack99},
\begin{equation}
\chi_{d}(\Delta_n)=\frac{1}{2}(\delta(d-\Delta_n)+\delta(d+\Delta_n)) \quad . \label{eq:drp}
\end{equation}
Since $|\Delta_n|\le d$, we denote $d$ as the perturbation strength. As an
example, we sketch in Fig.\ \ref{fig1} our model for noisy slopes.  We now
define the diffusion coefficient as
\begin{equation}
D(a,d)=\lim_{n\to\infty}\frac{1}{2n}(<x_n^2>_{\rho_0}-<x_n>^2_{\rho_0})\;, \label{eq:dkdef}
\end{equation}
with 
\begin{equation}
<x_n^k>_{\rho_0}=\int dx\int d(\Delta_0) d(\Delta_1) \ldots d(\Delta_{n-1})
\rho_0(x) \chi(\Delta_0)\chi(\Delta_1)\ldots\chi(\Delta_{n-1})x_n^k\;, \label{eq:dkdef2}
\end{equation}
where $\rho_0(x)$ denotes the initial distribution of an ensemble of moving
particles, $x_0\equiv x$, $k\in \hbox{\mb N}$, and $\Delta_j ,
j\in\left\{1,\ldots,n-1\right\}$, is the random variable. Note that in
computer simulations it suffices to generate a single series of random
variables instead of evaluating all the integrals in Eq.\
(\ref{eq:dkdef2}). To obtain better numerical convergence for noisy shifts the
current squared in Eq.\ (\ref{eq:dkdef}) was subtracted at any time step while
Eqs.\ (\ref{eq:urp}),(\ref{eq:drp}) imply that the long-time average over the
random variable $\Delta_n$ does not yield any bias. In Refs.\
\cite{RKD,RKdiss} it was shown that the unperturbed map Eq.\ (\ref{eq:eom})
exhibits normal diffusion if $a>2$, and the same was found recently by adding
a bias $b$ \cite{KlGr00}. Correspondingly, for the types of perturbations
defined above diffusion should always be normal if $(a-da)>2$, as was
confirmed in simulations. Hence, the central question is what happens to the
parameter-dependent diffusion coefficient $D(a,d)$ under variation of the two
control parameters $a$ and $d$ in case of the above two types of noise.

For $da=0$ it was shown that the unperturbed diffusion coefficient $D(a,0)$ is
a fractal function of the slope $a$ as a control parameter
\cite{RKD,RKdiss}, as is depicted again in Fig.\ \ref{fig1}. Included are
results from computer simulations for uniformly distributed noisy slopes at
different values of the perturbation strength $da$ \cite{cs}. As expected, the
fractal structure gradually smoothes out by increasing $da$. Qualitatively the
same result is obtained by applying noisy shifts \cite{tbp}. Fig.\ \ref{fig1}
may be compared to the corresponding result for {\em quenched} slopes Fig.\ 1
in Ref.\ \cite{RKla01a}. Apart from numerical uncertainties, there are clear
differences in the critical behavior close to the onset of diffusion. However,
for small enough perturbation strength and large enough $a$ the results look
qualitatively similar indicating that in this limit quenched and annealed
diffusion may be treated on the same footing.

More evidence for this statement is obtained from a trivial approximation for
the perturbed diffusion coefficient, which we motivate starting from
dichotomous noisy slopes. Naive reasoning suggests that, at arbitrary fixed
parameters $a$ and $da$, the {\em perturbed} diffusion coefficient $D(a,da)$
can be approximated by simply averaging over the {\em unperturbed} diffusion
coefficient $D(a,0)$ at respective values of the slopes $a-da$ and $a+da$
yielding $D_{\rm app}(a,da)=(D(a-da,0)+D(a+da,0))/2$. This heuristic argument
can be straightforwardly extended to the random distribution Eq.\
(\ref{eq:urp}) as well as to any other type of uncorrelated noise yielding the
generalized expression
\begin{equation}
D_{\rm app}(\mbox{\boldmath $p$},\mbox{\boldmath $d$})=\int
d(\mbox{\boldmath $\Delta$})\: \chi_{d}(\mbox{\boldmath
$\Delta$})D(\mbox{\boldmath $p$}+\mbox{\boldmath $\Delta$},0)\;.
\label{eq:dappg}
\end{equation}
Here $\mbox{\boldmath $p$}$ is a vector of control parameters such as
$\mbox{\boldmath $p$}=\left\{a,b\right\}$ in case of the map above,
$\mbox{\boldmath $d$}$ is the corresponding vector of perturbation strengths,
and $\mbox{\boldmath $\Delta$}$ is the vector of perturbations such as
$\mbox{\boldmath $\Delta$}=\left\{\Delta a,\Delta b\right\}$ for noisy shifts
and slopes. Further generalizations of this equation, for example, to
arbitrary moments as defined in Eq.\ (\ref{eq:dkdef2}), are straightforward.
Applying this formula to the case of quenched slopes discussed in Ref.\
\cite{RKla01a} reproduces the diffusion coefficient approximation Eq.\ (6)
therein, which was obtained in the limit of small perturbation strength. The
corresponding approximations for uniform noisy slopes are depicted in Fig.\
\ref{fig1} as lines. They show that even for the rather large
perturbation strength $da=1$ the agreement between theory and simulations is
excellent.  This confirms that, in the limit described above, quenched and
annealed disorder generating normal diffusion can indeed approximately be
treated in the same way.

\begin{figure}[b]
\begin{center}
\epsfxsize=6.5cm
\subfigure{\rotate[r]{\epsfbox{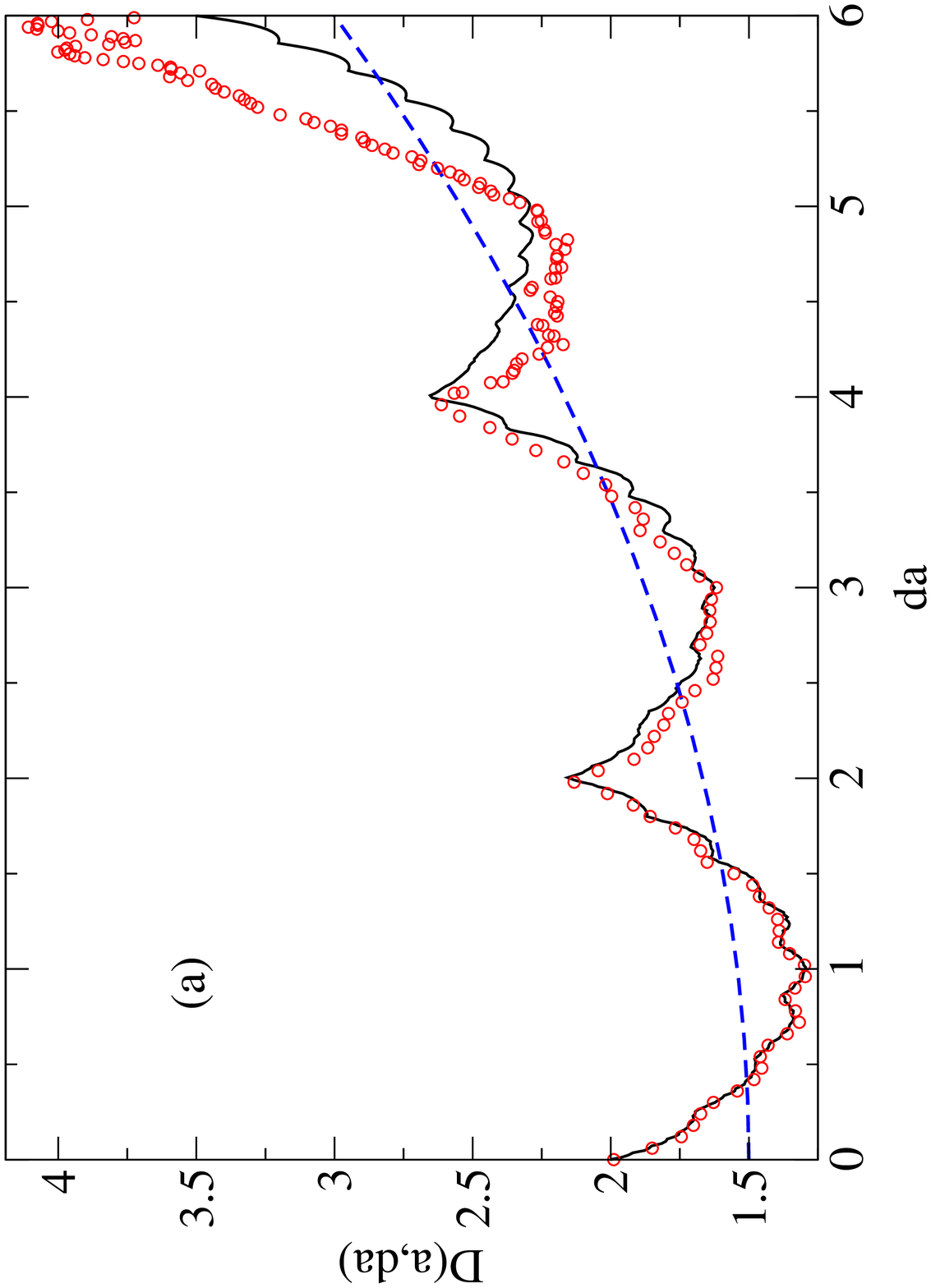}}}
\epsfxsize=6.5cm
\subfigure{\rotate[r]{\epsfbox{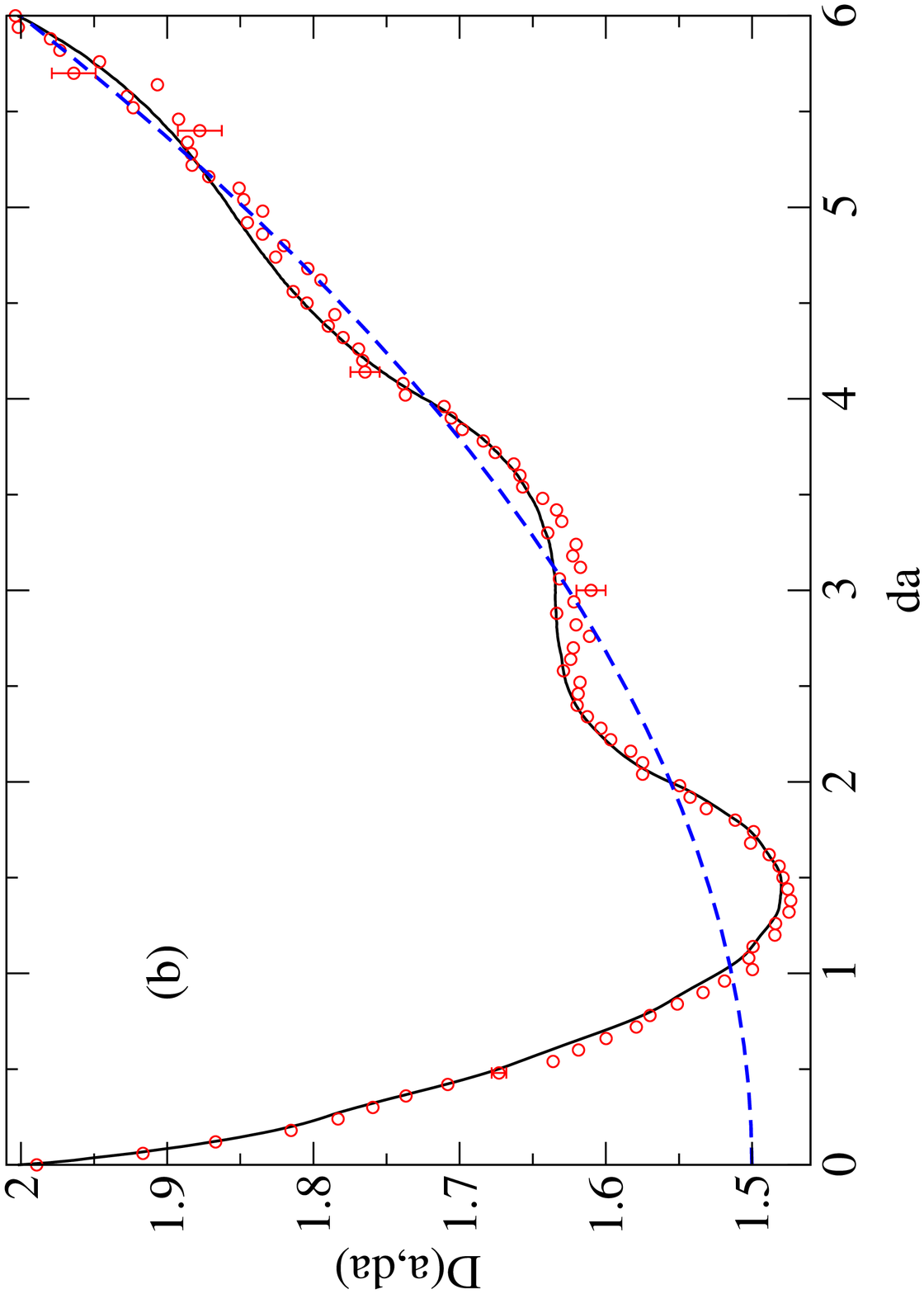}}}
\end{center}
\vspace*{-0.6cm}
\caption{Diffusion coefficient $D(a,da)$ as a function of the perturbation
strength $da$ at slope $a=7$ for noisy slopes distributed according to: (a)
dichotomous noise Eq.\ (\ref{eq:drp}), (b) uniform noise Eq.\
(\ref{eq:urp}). The circles represent results from computer simulations, the
bold lines are obtained from the approximation Eq.\ (\ref{eq:dappg}), the
dashed lines represent the stochastic limit for the diffusion coefficient Eq.\
(\ref{eq:drwp}).}
\label{fig2}
\end{figure}

Let us now look at the diffusion coefficient for a given value of $a$ as a
function of $da$. Fig.\ \ref{fig1} shows that approximately at odd and even
integer slopes the fractal diffusion coefficient $D(a,0)$ exhibits a local
maximum or minimum, respectively. Since Eq.\ (\ref{eq:dappg}) represents an
average over the unperturbed solution in a local environment $[a-da,a+da]$ it
predicts local suppression and enhancement of diffusion at odd and even
integer slopes, respectively, under variation of the perturbation strength
$da$. This has already been conjectured in Ref.\
\cite{RKdiss} and has been verified in Ref.\ \cite{RKla01a} for quenched
slopes. We first check this hypothesis for noisy slopes around the local
maximum of $D(a,0)$ at $a=7$ distributed according to Eqs.\
(\ref{eq:urp}),(\ref{eq:drp}). Figs.\ \ref{fig2} (a), (b) depict again results
obtained from computer simulations in comparison to Eq.\ (\ref{eq:dappg}). As
predicted, in both cases there is suppression of diffusion for small enough
$da$. For dichotomous noise the perturbed diffusion coefficient increases on a
coarse scale by exhibiting multiple, fractal-like suppression and enhancement
on finer scales. For uniform perturbations there is a pronounced crossover
from suppression to enhancement on a coarse scale, by again exhibiting
oscillations on a fine scale. In both cases the agreement between simple
theory and simulations is excellent for small enough $da$, whereas clear
systematic deviations particularly in case of dichotomous noise are visible
for larger $da$. Note that if $a-\Delta a_n<2$ particles are getting trapped
within a box at a respective time step $n$, and that for $a-\Delta a_n<1$ the
map is non-chaotic. In the first case simulations and simple reasoning suggest
that the perturbed map still exhibits normal diffusion. However, as soon as
$a-da<1$ numerical results indicate that there is no normal diffusion anymore
\cite{tbp}. This appears to be due to the contracting behavior of the
non-chaotic map resulting in localization of particles. The oscillatory
behavior of the diffusion coefficient in Fig.\ \ref{fig2} (a) just below this
transition point is not yet understood.

\begin{figure}
\begin{center}
\epsfxsize=6.5cm
\subfigure{\rotate[r]{\epsfbox{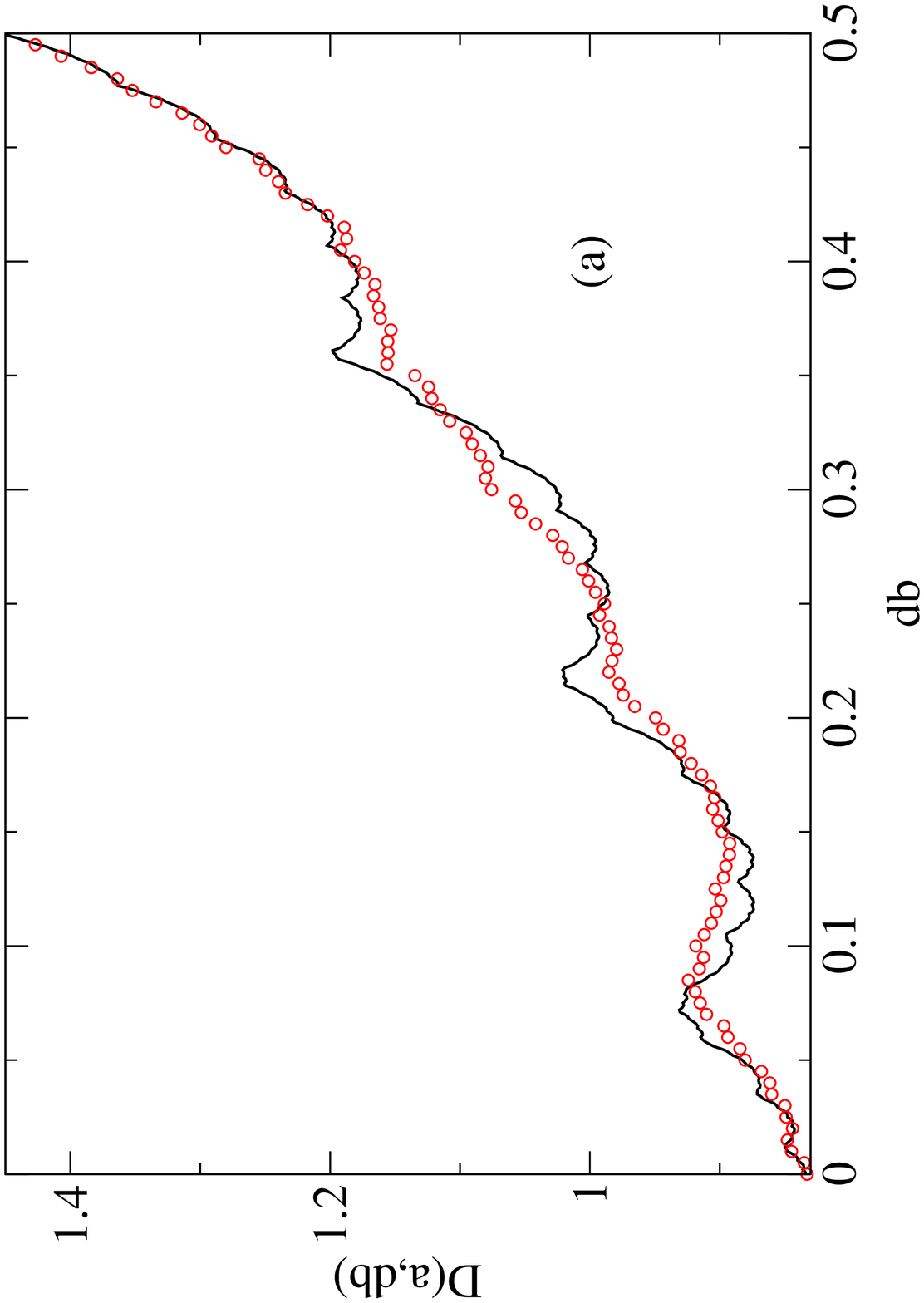}}}
\epsfxsize=6.5cm
\subfigure{\rotate[r]{\epsfbox{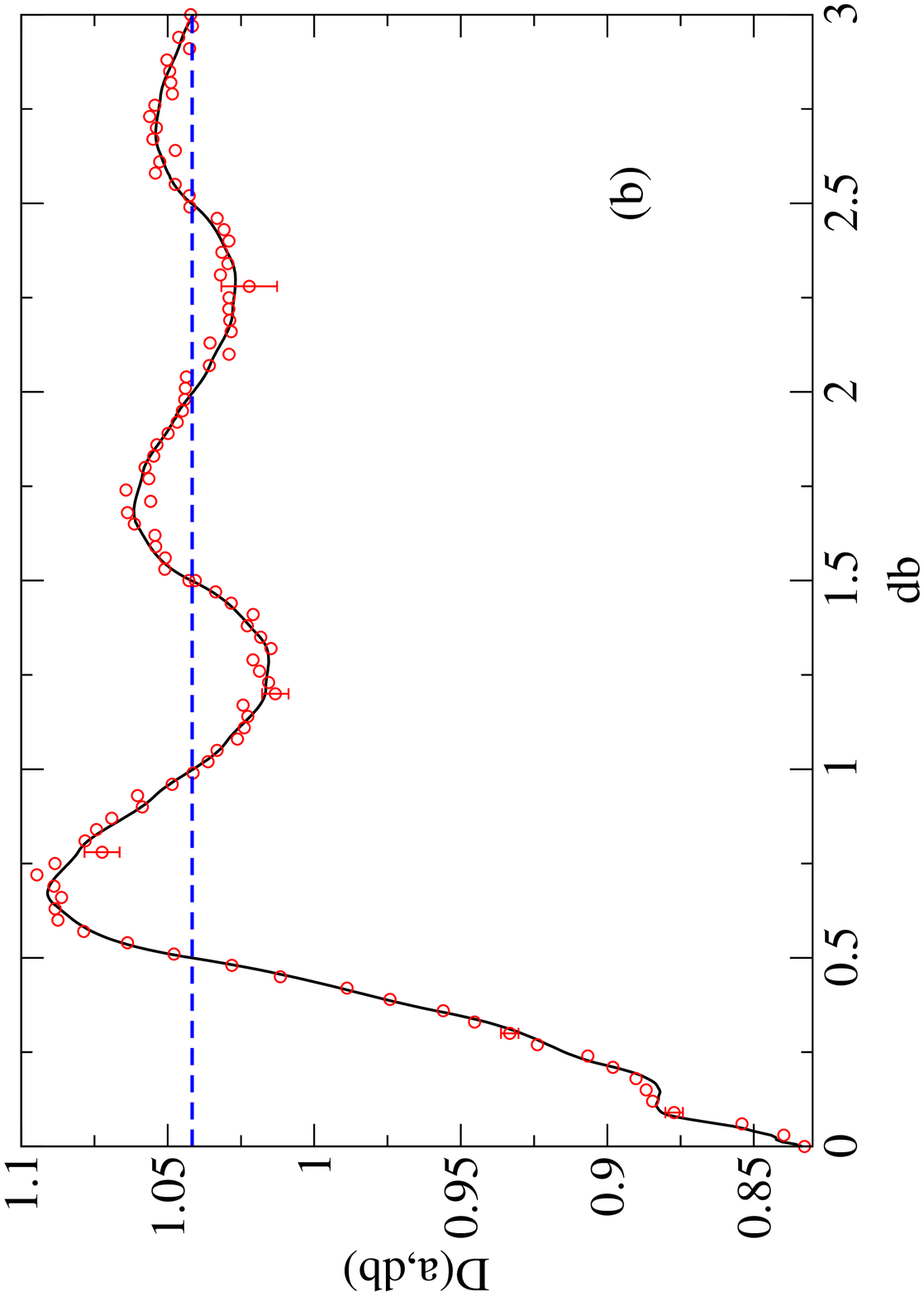}}}
\end{center}
\vspace*{-0.6cm}
\caption{Diffusion coefficient $D(a,db)$ as a function of the perturbation
strength $db$ at slope $a=6$ for noisy shifts distributed according to: (a)
dichotomous noise Eq.\ (\ref{eq:drp}), (b) uniform noise Eq.\
(\ref{eq:urp}). The circles represent results from computer simulations, the
bold lines are obtained from the approximation Eq.\ (\ref{eq:dappg}), the
dashed line represents the stochastic limit for the diffusion coefficient Eq.\
(\ref{eq:drw}).}
\label{fig3}
\end{figure}

Employing Eq.\ (\ref{eq:dappg}) we now analyze noisy shifts. The unperturbed
two-parameter diffusion coefficient $D(a,b,0)$ has been calculated numerically
exactly for the map under consideration in Ref.\ \cite{KlGr00}. Results for
the perturbed diffusion coefficient $D(a,db)\equiv D(a,0,db)$ are presented in
Fig.\ \ref{fig3} (a) for dichotomous noise and (b) for uniform perturbations,
both starting from $D(a,0)$ at $a=6$.  In both cases the perturbed diffusion
coefficient exhibits strong enhancement of diffusion for small enough
perturbation strength due to the fact that the unperturbed diffusion
coefficient at $a=6$ is approximately identical with a local maximum in the
$(a,b)$ parameter plane \cite{KlGr00}. For dichotmous perturbations it
suffices to show results for $0<db<0.5$ only. Translation and reflection
symmetry of the map imply that this function is mirrored in the interval from
$0.5<db<1$, and that the full sequence in $0<db<1$ is periodically repeated
for higher values of $db$.  As in the corresponding case of noisy slopes, the
perturbed diffusion coefficient increases on a coarse scale by exhibiting
multiple fractal-like suppression and enhancement on a fine scale. In case of
uniform perturbations there is a pronounced crossover to an approximately
constant diffusion coefficient for larger $db$.

Before calculating the stochastic limit of the diffusion coefficient we
provide a simple analytical justification for the heuristic approximation
Eq. (\ref{eq:dappg}). For sake of simplicity, we demonstrate it only for noisy
slopes, $\Delta_n\equiv\Delta a_n$.  Noisy shifts as well as quenched disorder
can be treated along the same lines \cite{tbp}. Let us start from the
definition of the diffusion coefficient Eq.\ (\ref{eq:dkdef}) where
$<x_n>=0$. Let $\Delta a_n$ be uniformly distributed in $[-da,da]$, $\Delta
a_0\equiv \Delta a$. In case of $da\to 0$ all random variables are bounded by
$\Delta a_n=\Delta a+\epsilon$, $-2da\le\epsilon\le2da$. We now put this
expression into the perturbed equation of motion Eqs.\
(\ref{eq:eom}),(\ref{eq:rs}), as contained in Eq.\ (\ref{eq:dkdef}), which we
write as $x_{n+1,a+\Delta a_n}=M_{a+\Delta a_n}(x_n)$.  As a first step we now
take the limit $\epsilon\to0$ resulting in the expression for the mean square
displacement
\begin{eqnarray} 
<x_n^2>&=&\int dx\int d(\Delta a) d(\Delta a_1) \ldots d(\Delta a_{n-1})
\rho_0(x) \chi(\Delta a)\chi(\Delta a_1)\ldots\chi(\Delta a_{n-1})
x^2_{n,a+\Delta a_{n-1}} \nonumber \\ &=& \int dx\int d(\Delta a) \rho_0(x)
\chi(\Delta a)x^2_{n,a+\Delta a}\;(\epsilon\to0)\;.
\end{eqnarray}
As a second step we exchange the time limit contained in Eq.\ (\ref{eq:dkdef})
with the integration over $d(\Delta a)$ yielding
\begin{eqnarray} 
D_{app}(a,da)&=&\lim_{n\to\infty}\frac{<x_n^2>}{2n}\nonumber \\
&=&\int d(\Delta a) \chi_{da}(\Delta a) \lim_{n\to\infty}\int dx
\rho_0(x)\frac{x^2_{n,a+\Delta a}}{2n} \nonumber \\ 
&=&\int d(\Delta a) \chi_{da}(\Delta a) D(a+\Delta a,0)\;, \label{eq:dkap}
\end{eqnarray}
where we have used that the unperturbed diffusion coefficient was defined as
\begin{equation}
D(a,0)=\lim_{n\to\infty}\int dx \rho_0(x) x_{n,a}^2\;. \label{eq:dka}
\end{equation}
We have thus verified our previous approximation Eq.\ (\ref{eq:dappg}) for
noisy slopes in the limit of small perturbation strength. A similar derivation
can be carried out for noisy shifts arriving again at Eq.\ (\ref{eq:dappg}) in
case of very small perturbation strength. For quenched shifts it is known that
a normal diffusion coefficient does not exist \cite{Rado96}, thus any
approximation by Eq.\ (\ref{eq:dappg}) must fail. Indeed, it turns out that in
this case taking the limit $\epsilon\to0$ fundamentally changes the properties
of the dynamical system and is thus no valid operation \cite{tbp}.

Finally, we calculate the parameter-dependent stochastic diffusion coefficient
related to the map with noisy slopes. Starting from the definition Eq.\
(\ref{eq:dka}) the complete loss of memory in the unperturbed map is modeled
by \cite{RKdiss,dcrc} (i) replacing the distance $x_n$ a particle travels by
$n$ times the distance a particle travels at any single time step, $n\Delta
x=n(M_a(x)-x)$, and (ii) neglecting any memory effects in the probability
density on the unit interval by assuming $\rho_0(x)=1$. Then Eq.\
(\ref{eq:dka}) yields
\begin{equation}
D_{rw}(a)=\frac{(a-1)^2}{24} \label{eq:drw}\;.
\end{equation}
As was shown in Refs.\ \cite{RKdiss,dcrc}, this equation correctly describes
the asymptotic parameter dependence of the deterministic diffusion coefficient
for $a\to\infty$ thus explaining the increase of $D(a,0)$ in Fig.\ \ref{fig1}
on a coarse scale.  On this basis, the corresponding result for noisy slopes
is easily calculated by using Eq.\ (\ref{eq:drw}) as the functional form for
$D(a+\Delta a,0)$ in the approximation Eq.\ (\ref{eq:dkap}) reading
\begin{equation}
D_{rw}(a,da)=D_{rw }(a,0)+\Delta a^2/c\;, \label{eq:drwp}
\end{equation}
where $c=24$ for dichotomous noise Eq.\ (\ref{eq:drp}) and $c=72$ for uniform
noise Eq.\ (\ref{eq:urp}). Eq.\ (\ref{eq:drwp}) thus confirms the common sense
expectation that noise should typically enhance diffusion and represents the
{\em stochastic limit} of the diffusion coefficient. This equation is depicted
in Fig.\ \ref{fig2} (a), (b) in form of dashed lines. In case of dichotomous
noise the correlations are apparently large enough such that even for large
perturbation strength $da$ there is no transition to the stochastic limit,
whereas in case of uniform noisy slopes the diffusion coefficient approaches
the stochastic solution asymptotically in $da$ thus verifying the existence of
a transition from deterministic to stochastic diffusion. That such a distinct
transition behavior exists in these models was already conjectured in Ref.\
\cite{RKdiss}. Analogous calculations for noisy shifts yield Eq.\
(\ref{eq:drw}) for all values of $db$ reflecting the fact that for large
enough $a$ the stochastic diffusion coefficient should not depend on the bias.
This result is shown in Fig.\ \ref{fig3} (b) and again confirms an asymptotic
approach of the diffusion coefficient to the stochastic limit under variation
of $db$. Based on the known result of the existence of a fractal diffusion
coefficient for the unperturbed $D(a,b,0)$ we conjecture that the typical
transition scenario in this type of systems consists of (multiple) suppression
and enhancement of diffusion. We finally note that Eqs.\
(\ref{eq:drw}),(\ref{eq:drwp}) are closely related to the approximation
outlined in Ref.\ \cite{GeNi82}, and to the simple heuristic argument given by
Reimann \cite{Rei94} by which he explains the suppression of deterministic
diffusion by noise in the climbing sine map near a crisis; more details will
be discussed elsewhere \cite{tbp}.

We conclude with a few remarks: (1) It would be interesting to study the
problem of noisy maps with non-zero average bias along the same lines. Ref.\
\cite{KlGr00} shows that the unperturbed map does not exhibit linear response
for $b\to0$, thus we conjecture that adding noise generates a transition to
Ohm's law.  (2) In the recent Ref.\ \cite{ViRu01} a rather general mechanism
of noise suppression by noise has been reported. Whether there is a more
detailed relation between the argument outlined in this reference and the
phenomena discussed here appears to be an open question. (3) Our approach may
be useful to investigate the impact of noise on the diffusion coefficient in
more complex time-continuous systems as well. In particular, we are thinking
of models such as the standard map, particle billiards, or inertia ratchets,
where irregular transport coefficients have already been reported and studied
under the impact of noise \cite{JuHa96}. However, these analyses were not
performed from the point of view of suppression and enhancement of diffusion,
or by looking for transitions to the stochastic limit.

The author thanks N.\ Korabel and S.\ Denisov for interesting discussions on
noisy maps.  He is also grateful to G. Radons, H.\ van Beijeren, J.R.\
Dorfman, and T.\ Tel for helpful remarks.\\[-4ex]


\begin{thebibliography}{10}
\bibitem{cs}
A maximum of 1,000,000 particles has been iterated up to 50,000 time steps
each. In figures without error bars, the numerical error is less than the size
of the symbols.
\end{thebibliography}

\end{document}